\documentstyle[epsf,twocolumn]{article}
\begin{document}

\begin{titlepage}
\begin{center}
{\Huge
Leptonic Decays of Mesons in a Poincare- Covariant of a
Quark Model
}
$$
$$
$$
$$

{\Large
V.V. Andreev\\
Gomel State University, Physics Department,
Gomel, 246699,Belarus.  \\
E-mail: andreev@gsu.unibel.by\\}
$$
$$
{\large
Published in Proceedings
7 Annual Seminar "Nonlinear Phenomena In Complex System" (NPCS'98)\\
( Minsk, Belarus, 1998  )\\ .}
\end{center}
$$
$$
$$
$$
\begin{center}
{\Huge Abstract}
\end{center}

\begin{center}

{\Large
Using a relativistic constituent quark model based on point form of
relativistic quantum mechanics with given set of degrees of  freedom
we investigate electroweak decays of the pion and $\rho$ mesons.
All free parameters of quark model (the mass u(d) -quark and
parameters which determines the confinement scale) are fixed by
using relevant experimental data.
}
\end{center}
\end{titlepage}
\newpage

{\Large
Leptonic Decays of Mesons in a Poincare- Covariant of a
Quark Model
}
$$
$$
{\large
V.V. Andreev\\
Gomel State University, Physics Department,
Gomel, 246699,Belarus.  \\
E-mail: andreev@gsu.unibel.by\\}

\section{\bf Introduction}

The researches of electroweak decays of hadrons always were a important of
an information about interaction of quarks. Today, electroweak decays of
hadrons, which contain heavy quarks enable to measure parameters of a
Standard Model (SM), and also serve for searches of effects of new physics
i.e. physics outside of SM. In particular, the hadronic decays allow to
define the elements of a matrix a Cabibbo-Kobayashi-Maskawa, angles of
mixing. The leptonic decays of pseudoscalar mesons in models with two
charged Higgs bosons become sensing to masses of these bosons \cite{izv1} .
Such researches require the registration of a structure of hadrons, in
particularly, of mesons. Now for exposition of relativistic bound systems
there is a set of modes. Within the framework of the given paper even the
brief enumeration would take a lot of places. In the given work for an
evaluation of leptonic constants of connection of mesons consisting from
quarks of equal mass, we use a relativistic quark model based on a point
form a Poincare - invariant of a quantum mechanics (other frequently used
title of this mechanics - relativistic Hamiltonian dynamics( see, for
example, \cite{izv1a} ~). The purpose of work consists in an evaluation of
constants of decays of mesons in the given approach with the help of model
wave function and fixing of free parameters of a model with the help of
experimental data. There are three forms of dynamics in a relativistic
quantum mechanics (hereinafter simple RQM): point both instant the form and
dynamics on light-front form ~ \cite{izv2} ~. At exposition of electroweak
properties of mesons dynamics on light front widely is used. So in work ~
\cite{izv3} ~, ~ \cite{izv4} ~ the constants of decays of mesons and baryons
paid off. The remaining forms of dynamics are used less often. Within the
framework of an instant form a RQM the leptonic decays were calculated only
in work ~ \cite{izv5} ~, ~ \cite{izv6}. But the relativistic impulse
approximation in an instant form and dynamics on light front automatically
breaks a Poincare - invariance of models. In a point form the RQM of such
violation does not happen and consequently we use it in our calculations.

\section{\bf a RQM a formalism for $Q\bar q $ of bound state system}

Dynamics of multiparticle systems with specific number of degree of freedoms
in framework a RQM is determined by generators of group the Poincare $\hat
{M} _ {\mu\nu} $ and $\hat {P} _ {\mu} $, which are dynamic variables. At a
construction of generators for a system from interacting particles begin
from generators of an appropriate system composed from noninteracting
particles ({\bf hereinafter we shall note such operators without " hat "}),
and then add interaction so that the obtained generators also satisfied to
commutation relations of group the Poincare. In relativistic case it is
necessary to add interaction $V $ more than in one generator to satisfy with
algebra of group the Poincare. The Dirac ~ \cite{izv2} ~ has shown, that
there is no unequivocal(unambiguous) separation of generators on a dynamic
set (generators containing interaction $U $) and a kinematics set. The
kinematics set can be connected to some subgroup of group the Poincare
usually called as group of stability or the kinematics subgroup ~ \cite{izv7}
~. In common, this subgroup contacts to a three-dimensional hypersurface,
which remains invariant under an operation a Poincare of transformations
from the given subgroup. So for example, the instant form a RQM can be
connected to a hypersurface $t $ = 0, and form of dynamics on light front
with a plane $t + Z $ = 0.

The operators of three-dimensional momentum of a system and angular momentum
do not contain in an instant form of dynamics interaction i.e. ${\hat {\vec
P}} $ = ${\vec P} $ and ${\hat {\vec J}} $ = ${\vec J} $ (${\hat {\vec J}} $
= (${\hat M} ^ {23} $, ${\hat M} ^ {31} $, ${\hat M} ^ {12} $)), while a
mass operator $M $ (or Hamiltonian $\hat P _ 0 = \sqrt {M ^ {2} + \vec P ^
{2}} $) and operator of a boost ${\hat {\vec N}} = ({\hat M} ^ {01}, {\hat M}
^ {02}, {\hat M} ^ {03}) $ include addenda with interaction. The description
in a point form implies that the operators ${\hat M} ^ {\mu\nu} $ are same
as for noninteracting particles i.e. ${\hat M} ^ {\mu\nu} = M ^ {\mu\nu} $,
and the terms with interaction are contained only in an operator $4 $ -
momentum ${\hat P} $. In dynamics on light front we shall enter + and -
components 4- vectors by a usual way:
\[
P ^ + = ( p ^ 0 + p ^ z) /\sqrt {2}, ~~~ p ^ - = (p ^ 0-p ^ z) /\sqrt {2}.
\]
We require that in the front form the operators ${\hat P}^+,{\hat P}^j, {\
\hat M}^{12},{\hat M}^{+-}, {\hat M}^{+j}$ $(j=1,2)$ are the same as the
corresponding free operators, and interaction terms may be present in the
operators ${\hat M}^{-j}$ and ${\hat P}^-$. As the RQM 4-momentum of a bound
system $P $ and total 4-momentum of particles $P _ {12} = p _ {1} + p _ {2} $
, component this system (system of free quarks follows from explained above,
in all forms of dynamics , for example) do not coincide i.e.
\begin{equation}
P \ne P _ {12} = p _ {1} + p _ {2}.
\end{equation}

The momentum $p _ {i} $ of particles of a relativistic system (quarks in a
meson) can be transformed in full $\vec {P _ {12}} $ and relative momentum $%
\vec k $ for selection of motion of a center mass system:

\begin{equation}
\overrightarrow{k}=\overrightarrow{p_1}+\frac{\overrightarrow{P_{12}}}{M_0}
\left( \frac{\left( \overrightarrow{P_{12}}\overrightarrow{p_1}\right) }{
\omega _{M_0}\left( \overrightarrow{P_{12}}\right) +M_0}+\omega _{m_1}\left(
\overrightarrow{p_1}\right) \right) ,  \label{veck}
\end{equation}
Where
\begin{eqnarray}
M_0 &=&\omega _{m_1}\left( \overrightarrow{k}\right) +\omega _{m_2}\left(
\overrightarrow{k}\right) ,  \nonumber \\
\omega _{m_1}\left( \overrightarrow{p_1}\right) &=&\sqrt{\overrightarrow{p_1
}^2+m_1^2}.  \label{mass}
\end{eqnarray}

The problem on eigenvalues for mass $Q \bar q $ system can expressed as: ~
\cite{izv1a} ~
\begin{equation}
\hat M \mid\Psi > = (\hat M _ {0} + \hat U) \mid \Psi > = M \mid \Psi >,
\end{equation}

The eigenfunctions in a point form of dynamics look like
\[
\left\langle \vec V_{12}\hskip 2pt\mu ,\left[ J\hskip 2ptk\right]
,(ls)\right. \left| \vec V\hskip 2pt\mu ,\left[ J\hskip 2ptM\right]
\right\rangle =
\]
\begin{equation}
=2\sqrt{1+\vec V^2}\hskip 2pt\delta _{JJ^{\prime }}\delta _{\mu \mu ^{\prime
}}\delta (\vec V-\vec V_{12})\Psi ^{J\mu }\left( k\hskip 2ptl\hskip %
2pts\right)   \label{wfp}
\end{equation}
With 4-velocities of a bound system $\vec V={\vec P}/{M}$ and noninteracting
system $\vec V_{12}={\vec P_{12}}/{M_0}$. \noindent
The vector $\mid \vec V_{12}\hskip 2pt\mu ,\left[ J\hskip 2ptk\right] ,(l%
\hskip 2pts)>$ is the eigenvector of operators $\vec V_{12},k,\vec J^2$
(angular momentum), $\hat \mu =\hat J_3$ and also operators $\vec L^2$, $%
\vec S^2$, where $\vec L$ and $\vec S$ relative orbital and full spin moment
accordingly.

This state vector noninteracting $Q\bar q$ system sets base of a
two-particle representation of group the Poincare and will be transformed in
appropriate way (look \cite{izv1a}). The vector $\mid \vec V\hskip 2pt\mu
,\left[ J\hskip 2ptM\right] >$ is vector of interacting $Q\bar q$ system and
derived base of an one-particle representation of a group the Poincare with
appropriate by the law of transformation by induced Poincare transformation.

Inserting a full set of basis condition of a free system Let's receive the
equation for $\Psi ^{J\mu }(k,l,s)$:
\[
\sum_{l^{\prime }s^{\prime }}\int\limits_0^\infty <k\hskip
2ptL\hskip 2pts\parallel U^J\parallel k^{\prime }\hskip 2ptL^{\prime }\hskip %
2pts^{\prime }>\Psi ^{J\mu }(k^{\prime }\hskip 2ptL^{\prime }\hskip %
2pts^{\prime }){k^{\prime }}^2dk^{\prime }+
\]
\begin{equation}
+M_0\Psi ^{J\mu }(k\hskip 2ptl\hskip 2pts)=\lambda \Psi ^{J\mu }(k\hskip 2ptl%
\hskip 2pts)  \label{eqf}
\end{equation}
in a point form of dynamics a RQM with a reduced matrix operator $\hat U$.
As vector of irreducible basis for a system with interaction and without it
satisfy to conditions of a normalization and completeness, the wave
functions satisfy to a condition of a normalization:
\begin{equation}
\sum_{ls}\int_0^\infty dk\hskip 2ptk^2\hskip 2ptN_c\left| \Psi ^{J\mu
}\left( kls\right) \right| ^2=1.  \label{norm}
\end{equation}
Here colour degrees of freedom of quarks we take into account by
introduction of number of colours $N_c$.

In such approach state vector of a meson is determined as a direct product
of state vectors of free quarks with wave function $\Psi ^ {J\mu} \left (k %
\hskip 2pt \hskip 2pt s\right) $

\[
\left| \overrightarrow{V},\mu \hskip 2pt\left[ JM\right] \right\rangle =
\]
\[
=\sum_{ls}\sum_{\lambda _1\lambda _2}\int d^3k\sqrt{\frac{\omega
_{m_1}\left( \overrightarrow{p_1}\right) \omega _{m_2}\left( \overrightarrow{
P_2}\right) }{\omega _{m_1}\left( \overrightarrow{k}\right) \omega
_{m_2}\left( \overrightarrow{k}\right) }}(\sqrt{2}M_0^{3/2})
\]
\[
\ast \Psi ^{J\mu }\left( kls;M\right) \sum_{m\lambda }\sum_{\nu _1\nu
_2}\left\langle s_1\nu _1,s_2\nu _2\right| \left. S\lambda \right\rangle
\left\langle lm,s\lambda \right| \left. J\mu \right\rangle
\]
\[
\ast Y_{lm}\left( \theta ,\phi \right) D_{\lambda _1\nu _1}^{1/2}\left(
\overrightarrow{n}\left( p_1,P\right) \right) D_{\lambda _2\nu
_2}^{1/2}\left( \overrightarrow{n}\left( p_2,P\right) \right) *
\]
\begin{equation}
\left| p_1,\lambda _1\right\rangle \left| p_2,\lambda _2\right\rangle ,
\label{state}
\end{equation}
Where $\left\langle s_1\nu _1,s_2\nu _2\right| \left. S\lambda \right\rangle
$, $\left\langle lm,s\lambda \right| \left. J\mu \right\rangle $ -
Clebsch-Gordon coefficients for $SU(2)$ - group, $Y_{lm}(\theta ,\phi )$ -
surface harmonics with spherical angles of a vector $\vec k$. Also in the
equation (\ref{state}) $D^{1/2}\left( \overrightarrow{n}\right) =1-i\left(
\overrightarrow{n}\overrightarrow{\sigma }\right) /\sqrt{1+\overrightarrow{n}
^2}$ there is $D$ - a function of Wigner rotation, which is determined with
the help of vector - parameter $\overrightarrow{n}\left( p_1,p_2\right) =
\overrightarrow{u_1}\times \overrightarrow{u_2}/(1-\left( \overrightarrow{%
u_1 }\overrightarrow{u_2}\right) )$ and $\overrightarrow{u}=\overrightarrow{p%
} /\left( \omega _M\left( \overrightarrow{p}\right) +m\right) $.

\section{\bf The Basic requirements for an operator of a current}

The operators of a current $\hat J^{(}x))$ bound system are necessary for an
evaluation of constants of decays, charge form factors and other properties
of relativistic particles. As $\hat J^{(}x))$ is four-vector, its property
under an operation of transformations the Poincare same as well as at an
operator four - momentum $\hat P_\mu $. It is reduced in that commutation
relation between $\hat J^{(}x))$ and generators of group with the Poincare $%
\hat M^{\rho \sigma }$, $\hat P_\mu $ are identical to commutation relations
between $4$ - momentum and generators:
\begin{eqnarray}
\left[ \hat M^{\rho \sigma },\hat J^\mu (x)\right] &=&I\left( g^{\mu \sigma
}\hat J^\rho \left( x\right) -g^{\mu \rho }\hat J^\sigma \left( x\right)
\right) -  \nonumber \\
&&\ i\left( x^\rho \frac{\partial \hat J^\mu \left( x\right) }{\partial
x_\sigma }-x^\sigma \frac{\partial \hat J^\mu \left( x\right) }{\partial
x_\rho }\right) ,  \label{tok3}
\end{eqnarray}
\begin{equation}
\left[ \hat P_\mu ,\hat J_\gamma (x)\right] =-i\frac{\partial \hat J_\mu
\left( x\right) }{\partial x^\gamma }.  \label{tok4}
\end{equation}

The translation invariance for an operator of a current reduces in the
equation of an aspect:
\begin{equation}
\hat J_\mu \left( x\right) =\exp \left( i\hat PX\right) \hat J_\mu \left(
0\right) \exp \left( -i\hat Px\right) .  \label{tok5}
\end{equation}
This equation makes possible to reduce a problem of searching $\hat J_\mu
\left( x\right) $ to a problem of searching $\hat J_\mu (0)$. The
requirement of Lorentz invariance for $\hat J_\mu (0)$ reduces in a relation
\begin{equation}
\left[ \hat M^{\rho \sigma },\hat J^\mu (0)\right] =i\left( g^{\mu \sigma
}\hat J^\rho \left( 0\right) -g^{\mu \rho }\hat J^\sigma \left( 0\right)
\right) .  \label{tok6}
\end{equation}
If the theory is invariant in relation to spatial inversion and reflection
of time, and the operators ${\hat U}_P$, ${\hat U}_T$ are operators of these
transformations, the operator of a current should satisfy to following
conditions:
\[
\widehat{U}_P({\hat J}^0(x^0,{\vec x}),{\vec J}(x^0,{\vec x})){\hat U}
_P^{-1}=
\]

\[
=({\hat J}^0(x^0,-{\vec x}),-{\vec J}(x^0,-{\vec x})),
\]
\[
\widehat{U}_T({\hat J}^0(x^0,{\vec x}),{\vec J}(x^0,{\vec x})){\hat U}
_T^{-1}=
\]
\begin{equation}
=({\hat J}^0(-x^0,{\vec x}),-{\vec J}(-x^0,{\vec x}))  \label{ptime}
\end{equation}
In addition to these equations the law can be used conservation of a current
(equation of a continuity) $\partial \hat J^\mu \left( x\right) /\partial
x^\mu $ $=0$. As follows from (\ref{tok4}) this requirement can be noted in
to the form:
\begin{equation}
\left[ \hat P_\mu ,\hat J_\nu (0)\right] g^{\mu \nu }=0.  \label{tok8}
\end{equation}
At last, the operator $\hat J_\mu \left( x\right) $ in a RQM should satisfy
to conditions of a so-called cluster separability (\cite{izv1a} ~) for
multiparticle systems.

At evaluations many authors suppose, that the mathematical expressions for
an operator of a current of a bound system and noninteracting system are
equal. This condition (so-called relativistic impulse approximation)
\begin{equation}  \label{tok7}
\hat {J} _ \mu \left (0\right) = J _ \mu (0)
\end{equation}
the Poincare - invariant of a relativistic quantum mechanics (~ \cite{izv8}
~) can be realized without any assumptions only in {\bf point} form of RQM.
This result follows from a relation (\ref{tok6}). In an instant form of
dynamics and in dynamics on light front the relativistic impulse
approximation (\ref{tok7}) automatically is reduced in violations of
commutation relations of Poincare group the for an current operator.

\section{\bf Leptonic constants of decays of mesons in a formalism a RQM}

Constant $f_p$ of a leptonic decay $P(Q\bar q)\to l+\nu _l$ for a
pseudoscalar meson $P(Q\bar q)$ after deleting the element of a matrix $%
V_{Qq}$, are usually determined by a following relation:
\begin{equation}
\left\langle 0\left| \hat J^\mu \left( 0\right) \right| \overrightarrow{P}
,M_P,in\right\rangle =i\left( 1/2\pi \right) ^{3/2}\frac 1{\sqrt{2\omega
_{M_P}\left( \overrightarrow{P}\right) }}P^\mu f_p,  \label{dec2}
\end{equation}
Where $\hat J^\mu (0)$, and state vector of a meson with mass $M_P$
undertake in Representation of the Heisenberg. State vector in this
expression has Normalization: $\left\langle \vec P,M\right. \left| \vec
P^{\prime },M\right\rangle $ = $\delta (\vec P-\vec P^{\prime })$.

The extraction of a constant of a decay from the matrix element of a current
(\ref{dec2}) is an independent problem and consequently we stay on it more
in detail. For it we shall copy expression (\ref{dec2}) as follows:
\[
j_\mu ^P\equiv
\]
\[
\equiv \left\langle 0\right| T\{J_\mu ^H(0)\exp [i\int H_{tot}^h(x)dx]\}*{%
f(\hat M)}\left| \overrightarrow{V},M_P\right\rangle =
\]
\begin{equation}
=i(1/2\pi )^{3/2}f_PV_\mu ,  \label{iz2}
\end{equation}
where $F(M)=1/M^2$, $H_{tot}^h(x)=H_0^h(x)+H_{int}^h(x)$ - total Hamiltonian
interactions of quarks in a meson, and state vector of a meson with a
velocity $V$ is normalized with as follows:
\begin{equation}
\left\langle \overrightarrow{V},M\right. \left| \overrightarrow{V}%
,M\right\rangle =2V_0\delta \left( \overrightarrow{V}-\overrightarrow{%
V^{\prime }}\right) ,V_0=\omega _M\left( \overrightarrow{P}\right) /M.
\label{iz2b}
\end{equation}
Further in expression (\ref{iz2b}) inserting a full set of state vectors of
two-particle system, forming base of an irreducible representation of
Poincare group for free particles and using a Lippmann-Schwinger equation
\begin{equation}
\left| p,in\right\rangle =\left( 1+\lim\limits_{\varepsilon \longrightarrow
+0}\frac 1{M-M_0+i\varepsilon }U\right) \left| p\right\rangle _V,
\label{iz1}
\end{equation}
( for a simplicity we have noted it in a system of rest of a meson, as all
calculations we shall carry out in this frame of reference), which links
state vector with representation of the Heisenberg and in representation
interactions (state vector with an index $V$), we come to the formula for
the matrix element of a current:
\[
j_\mu ^P=
\]
\[
=\int k^2dk\left\langle 0\right| T\{J_\mu ^H(0)\exp [i\int
H_{int}^h(x)dx]\}\left| \overrightarrow{V_{12}},k\right\rangle _V*
\]
\begin{equation}
\ast \psi ^P\left( k\right) (1/M_0^2)  \label{iz4}
\end{equation}
with wave function for a pseudoscalar meson $\psi ^P\left( k\right) $ ( see
(\ref{wfp})). Thus we also took into account, that owing to the
Lippmann-Schwinger equation the relation is fulfilled:
\begin{equation}
\hat M\left| \overrightarrow{V}_{12},k\right\rangle _V=M_0\left|
\overrightarrow{V}_{12},k\right\rangle _V.  \label{iz2a}
\end{equation}
Electroweak current of a free two-quark system in the formula ( \ref{iz4})
is determined by a form factor $G(k)$ i.e.

\begin{eqnarray}
&&\left\langle 0\right| T\{J_\mu ^H(0)\exp [i\int H_{int}^h(x)dx]\}\left|
\overrightarrow{V_{12}},k\right\rangle _V(1/M_0^2)  \nonumber \\
&=&i(1/2\pi )^{3/2}G_0(k)V_{12\mu }.  \label{iz5}
\end{eqnarray}
An explicit of the matrix element (\ref{iz5}) follows from method
parametrization of the matrix elements of local operators . As in a point
form of a vector of 4-velocities meson $V_\mu $ and free system $V_{12\mu }$
coincide, we shall receive, that the constant of a leptonic decay can be
calculated from expression:
\begin{equation}
f_p=\int\limits_0^\infty k^2\hskip 2ptdk\psi ^P(k)G_0(k).  \label{iz6}
\end{equation}
The form factor $G_0(k)$ can be calculated within the framework of a
Standard Model by expansion (\ref{iz5}) on base of a direct product of
one-particle state of quarks $\left| p_1,\lambda _1,p_2,\lambda
_2\right\rangle $. The equality (\ref{iz5}) can be put in a form:
\[
i(1/2\pi )^{3/2}G_0(k)V_{12\mu }=
\]
\[
=\sum \int \frac{d^3p_1d^3p_2}{2\omega _{m_1}\left( \overrightarrow{p_1}%
\right) 2\omega _{m_2}\left( \overrightarrow{p_2}\right) }*
\]
\[
\left\langle 0\right| T\{J_\mu ^H(0)\exp [i\int H_{int}^h(x)dx]\}\left|
p_1,\lambda _1,p_2,\lambda _2\right\rangle _V*
\]
\begin{equation}
\ast \left\langle p_1,\lambda _1,p_2,\lambda _2\right| \left|
\overrightarrow{V_{12}},k\right\rangle (1/M_0^2).  \label{iz7}
\end{equation}
In expression (\ref{iz7}) operator of a current of a meson $J_\mu ^H(0)$ and
Hamiltonian strong interaction $H_{int}^h(0)$ in a point form of dynamics
the RQM without any assumptions can be defined through operators of free
quark fields. Thus, as we marked early, the Poincare - invariance of a model
in an outcome of this approximation is not breaker, in difference from an
instant form of dynamics and dynamics on light front. Let's mark also, that
the wave function $\psi ^P(k)$ is Lorentz-invariant, as in the point form of
dynamics the operator of a boost does not contain interaction. In an instant
form a RQM this property of wave functions has not a place.

The expression for Clebsch-Gordon coefficients of Poincare group is known (
look, for example, \cite{izv1a} ), and matrix element of a current (\ref{iz7}
) in quark base in case of a leptonic decay by an operator of a current $%
\hat {J} ^ {H} _ {\mu} \left (0\right) = {\hat{\bar d}} \left
(0\right)
\gamma ^ \mu (1-\gamma _ 5) {\hat u} \left( 0\right) $ is given by standard
expression:

\[
\left\langle 0\right| J_\mu ^H(0)\left| p_1,\lambda _1,p_2,\lambda
_2\right\rangle _V=
\]
\begin{equation}
=1/(2\pi )^3\overline{V}^{\lambda _2}(\overrightarrow{p_2},m_2,)\gamma _\mu
(1-\gamma _5)U^{\lambda _1}(\overrightarrow{p_1},m_1,)  \label{iz8}
\end{equation}
The similar scheme can be realized not only for decays of pseudoscalar
mesons, but also for decays of vector mesons, and also for other reaction
including of mesons. After an integration in a system of rest of a meson, we
obtain following expression for leptonic constant of connection of a pion $%
f_\pi $ in the supposition of equality masses $u$ and $d$ of quarks ($%
m_u=m_d=m$) and ignoring QCD corrections :
\begin{equation}
F_\pi =\frac{N_cm}\pi \int_0^\infty \frac{k^2}{\omega _m^{3/2}\left(
\overrightarrow{k}\right) }\psi ^P(k)dk  \label{dec1}
\end{equation}
Let's remark, that as the calculations were carried out in a system of rest
of a meson, where the point form a RQM coincides an instant form, the
outcome (\ref{dec1}) coincides an outcome for this constants obtained in
\cite{izv6} ~ within the framework of an instant form of dynamics .

The constant of a decay $f_V$ is determined from the matrix element of
current
\[
\left\langle 0\right| J_\alpha ^H(0)\left| \overrightarrow{P},M_V,J=1,\mu
\right\rangle =
\]
\begin{equation}
=(1/2\pi )^{3/2}f_V\frac{M_V}{\sqrt{2\omega _{M_V}\left( \overrightarrow{P}
\right) }}\varepsilon _\alpha ^\mu \left( p\right) ,  \label{iz9}
\end{equation}
Where $\varepsilon _\alpha ^\mu \left( p\right) $ - polarization vector
meson with mass $M_V$. In (\ref{iz9}) the meson current can be selected by
an electromagnetic current of quarks
\begin{equation}
J_\alpha ^H(0)=e_u\overline{u}\gamma _\alpha u+e_d\overline{d}\gamma _\alpha
d  \label{iz10}
\end{equation}
for a decay $\rho ^0=(u\bar u-d\bar d)/\sqrt{2}$ - meson. Following
similarly to case of a decay of a pseudoscalar meson we shall receive:
\begin{equation}
F_\rho =\frac{N_c}{3\sqrt{2}\pi }\int_0^\infty \frac{k^2}{\sqrt{\omega
_M\left( \overrightarrow{k}\right) }}dk\left( 2+\frac m{\omega _m\left(
\overrightarrow{k}\right) }\right) \psi ^P(k)  \label{iz11}
\end{equation}
This expression is calculated for a longitudinally polarized condition $\rho
^0$ of a meson in a neglect by spin effects in quark-antiquark interaction.
In common case, wave functions in the formulas for $f_\pi $ and $f_\rho $
should differ.

The equation of motion for bound $Q\bar q$ systems (\ref{eqf}) in a RQM is
the relativistic equation with an effective potential $U$. However deriving
of wave functions $\psi (k)$ and spectrum of masses from this equation is a
challenge, which requires a separate research. In the given work we use
simple model wave function, which depends on a scale parameter $1/\beta $:
\begin{equation}
\Psi _P(k)\equiv \Psi _P(k\hskip 2pt,\beta )=2/(\sqrt{N_c}\beta ^{3/2}\pi
^{1/4})Exp(-\frac{k^2}{2\beta ^2}).  \label{wavef}
\end{equation}
Thus, in the given approach for pseudoscalar and vector mesons we have two
free parameters is a mass $u(d)$ of a quark $M$ and parameter $\beta $.
Expressions for leptonic constants with wave function (\ref{wavef}) can be
obtained in an analytical form with the help of special functions. So for a
pion we have:
\begin{eqnarray}
f_\pi  &=&\frac{\sqrt{3}m}{\pi ^{5/4}\Gamma \left( -\frac 14\right) }*
\nonumber \\
&&\ast (2^{3/4}\Gamma \left( -\frac 14\right) \Gamma \left( \frac 34\right) {%
_1F_1}\left( \frac 34;\frac 14;\frac{m^2}{2\beta ^2}\right) -  \nonumber \\
&&-\frac{2m^{3/2}\sqrt{\pi }}{\beta ^{3/2}}\Gamma \left( -\frac 34\right) {%
_1F_1}\left( \frac 32;\frac 74;\frac{m^2}{2\beta ^2}\right) ),  \label{izv13}
\end{eqnarray}
where $_1F_1(a;b;z)$ - degenerate hypergeometric function, and $\Gamma (z)$
- gamma-function. Similar, but more bulky The relation is received and for
magnitude $f_\rho $. Parameters a Poincare - covariant of a quark model of
electroweak decays it is possible to fix using, appropriate experimental
data. In our calculations we used the following values for leptonic
constants of decays: $f_\pi =130.7\pm 0.46$ ~~ MeV ~ and $f_\rho =152.9\pm
4.8$ ~~ MeV ( constant of a decay $\rho ^0$ of a meson in a pair $e^{+}e^{-}$
) \cite{izv11}. In the total we shall receive admissible with a point of
view of an existing experimental data the following solved areas for mass of
a quark $m$ and parameter $\beta $ of wave function ~ (\ref{wavef}), which
satisfactorily describe both experiments, researched by us,:
\begin{equation}
m=247\pm 9~MeV,~~~~\beta =323\pm 8~~~MeV.
\end{equation}
The data intervals for mass of easy quarks and for a parameter $\beta $ will
be agreed restrictions obtained within the framework of the front form a RQM
(see \cite{izv3}) and a model, based on a solution of the equation of the
Salpeter (see \cite{izv12}). Using obtained parameters, we have a
possibility to calculate other, independent from considered by us,
interaction of mesons with light quarks.

The work is carried out at support of Byelorussian Republican Fund of
Fundamental Researches ( project N 96-326-17.02.96).

\end{document}